\documentclass[10pt,twocolumn]{revtex4-1}
\usepackage{graphicx}
\usepackage{epstopdf}
\usepackage{amsmath}
\usepackage{amssymb}
\usepackage{color}
\begin{document}

\title{Coulomb focusing in retrapped ionization with near-circularly polarized laser field}

\author{Xiang Huang$^{1}$}
\author{Qingbin Zhang$^{1}$}
\email{zhangqingbin@hust.edu.cn}
\author{Shengliang Xu$^{1}$}
\author{Xianglong Fu$^{1}$}
\author{Xu Han$^{1}$}
\author{Wei Cao$^{1}$}
\email{weicao@hust.edu.cn}
\author{Peixiang Lu$^{1,2}$}

\affiliation{%
$^1$Wuhan National Laboratory for Optoelectronics and School of Physics, Huazhong University of Science and Technology, Wuhan 430074, China\\
$^2$Laboratory of Optical Information Technology, Wuhan Institute of Technology, Wuhan 430205, China
}%

\date{\today}

\begin{abstract}
  The full three-dimensional photoelectron momentum distributions of argon are measured in intense near-circularly polarized laser fields. We observed that the transverse momentum distribution of ejected electrons by 410-nm near-circularly polarized field is unexpectedly narrowed with increasing laser intensity, which is contrary to the conventional rules predicted by adiabatic theory. By analyzing the momentum-resolved angular momentum distribution measured experimentally and the corresponding trajectories of ejected electrons semiclassically, the narrowing can be attributed to a temporary trapping and thereby focusing of a photoelectron by the atomic potential in a quasibound state. With the near-circularly polarized laser field, the strong Coulomb interaction with the rescattering electrons is avoided, thus the Coulomb focusing in the retrapped process is highlighted.  
  We believe that these findings will facilitate understanding and steering electron dynamics in the Coulomb coupled system.
\end{abstract}                         
\maketitle

In the last three decades, the interaction of atoms and molecules with intense laser fields has attracted considerable attention in forefront physics. The study on it has opened a large variety of interesting phenomena and applications. One of them is strong-field ionization. Atoms or molecules can be ionized with releasing an electron wave packet driven by a strong laser field. The ionized electron wave packets are responsible for many nonlinear strong-field phenomena, such as high-order harmonic generation (HHG) [1-4], above-threshold ionization (ATI) [5], and non-sequential double-ionization (NSDI) [6-9].

The early investigations related to strong-field ionization are usually treated with the three-step model neglecting the potential of the ion [10-13]. Despite their success, it is difficult for it to explain some nonperturbative effects: double-peak structure in the momentum distribution [14,15], frustrated tunneling ionization (FTI) [16-22], and multiphoton assisted recombination [23]. These effects are caused by the interaction between the liberated electron and the ion. Besides, it is well accepted that the Coulomb effect on the liberated electron is responsible for the low-energy structure (LES) [24-26]. The multiple forward scattering by the atomic core leads to this characteristic spikelike structure in the energy distribution of electrons emitted along the polarization direction. Another important consequence of the Coulomb effect is Coulomb focusing [27-29], which has an important impact on strong-field ionization, especially on the significant enhancement of NSDI [30-32].

To date, most theoretical and experimental studies of the Coulomb effect focus on the driving field of linear polarization. Recently, the elliptically polarized laser field is used to provide a rotating electric field within one laser cycle. It adds more dimensions to facilitate the study of the strong-field ionization. For instance, with near-circularly polarized laser, the Coulomb potential is found to directly cause the deflection of the outgoing electrons, and leads to an offset angle of the photoelectron distribution. The indispensable role of the Coulomb interplay indicates the necessity of further calibration for the accurate reconstruction of ionization times [33]. This mechanism has been widely used to study attosecond-resolved electron dynamics, such as the precise measurements of tunneling time delay [33-36] and the position of tunneling exit [37]. Meanwhile, circularly or near-circularly polarized pulses has been also used to produce spin-polarized electrons at ultraviolet wavelengths. With the circularly polarized light at 400 nm, the degree of photoelectron spin polarization can reach as high as $\thicksim90\%$ [38], which provides a new direction of investigating polarized low-energy electron diffraction [39], probing the magnetic properties of condensed matters [40], and the source of polarized electron accelerators. In this relatively short wavelength range, however, the photoelectron is produced with lower energy, and will be more strongly affected by the Coulomb potential. In addition, a stronger nonadiabatic effect will be induced. There thus exists a lack of deep investigations on the photoelectrons from near-circularly or circularly polarized fields at a short wavelength. And the electron dynamic associated with Coulomb interaction in the ionization process is still confusing.

In this paper, we measure the full three-dimensional photoelectron momentum distributions (PMDs) of argon ionized by near-circularly polarized fields. The PMDs in the polarization plane by 410-nm near-circularly polarized field exhibit an abnormal spreading of the first-order ATI structure. And a remarkable energy-dependent angular shift is shown in the low-energy region. In the direction perpendicular to the laser field, the momentum distribution from near-circularly polarized field at 410 nm is unexpectedly narrowed while it is broadened at 800 nm with increasing field strength. This phenomenon clearly differs from the conventional rules predicted by tunneling theory. We reproduce the measurements by simulating the evolution of the emitted electron wave packet based on the nonadiabatic classical trajectory Monte Carlo approach (CTMC). Through disentangling different photoemission pathways, it is figured out that the photoelectron at the low-energy region will be temporarily retrapped by atomic potential into a quasibound state. The simulated results indicate the important role of Coulomb potential during retrapped ionization, which leads to retrapping and thereby significant focusing of the photoelectrons. These dynamics then result in the angle-shift and the spreading of first-order ATI structure to low-energy region in the polarization plane, and the narrowing distributions in the transverse direction.

We use a velocity map imaging (VMI) spectrometer to measure the three-dimensional momentum of photoelectrons [41]. The laser pulse ($\thicksim$35 fs, 800 nm) used in our experiments are generated from a Ti:sapphire femtosecond laser system with a repetition rate of 1 kHz. The combination of a $\lambda/2$ plate of 800 nm and a broadband wire grid polarizer is used to control the laser intensity. We use a $\lambda/2$ plate and a $\lambda/4$ of 800 nm to adjust the ellipticity of the laser field. Then we use a 1-mm-thick $\beta$-barium-borate crystal interacting with the near-infrared laser pulse to generate the 410nm laser pulse. The central wavelength is slightly detuned from 400 nm in order to benefit the generation of resonant ionization [42]. Early studies suggest that an excited atomic level can shift into resonance with the ground state, even upwards almost as much as the continuum level if the excited state is a high-lying Rydberg state. In this case, the energy of the first-order ATI peak will keep unchanged for varying laser intensity in a certain range [43-45]. With the intensity-independent peak positions, the first-order ATI peak can serve as a reliable reference to analyze the photoelectron with lower energy as the discussion below. The pulse duration at 410 nm is $\thicksim$70 fs. Elliptically polarized light is produced by a pair of $\lambda/2$ plate and $\lambda/4$ plates. Final full three-dimensional photoelectron momentum distributions are obtained by tomographic reconstruction [46]. To this end, the laser pulse is supposed to be rotated about its propagation direction with a $\lambda/2$ plate and a set of PMDs are recorded at different rotation angles. The rotation of the $\lambda/2$ plate is precisely controlled with a motorized rotator.

\begin{figure}
  \centering
  \includegraphics[width=8.5cm]{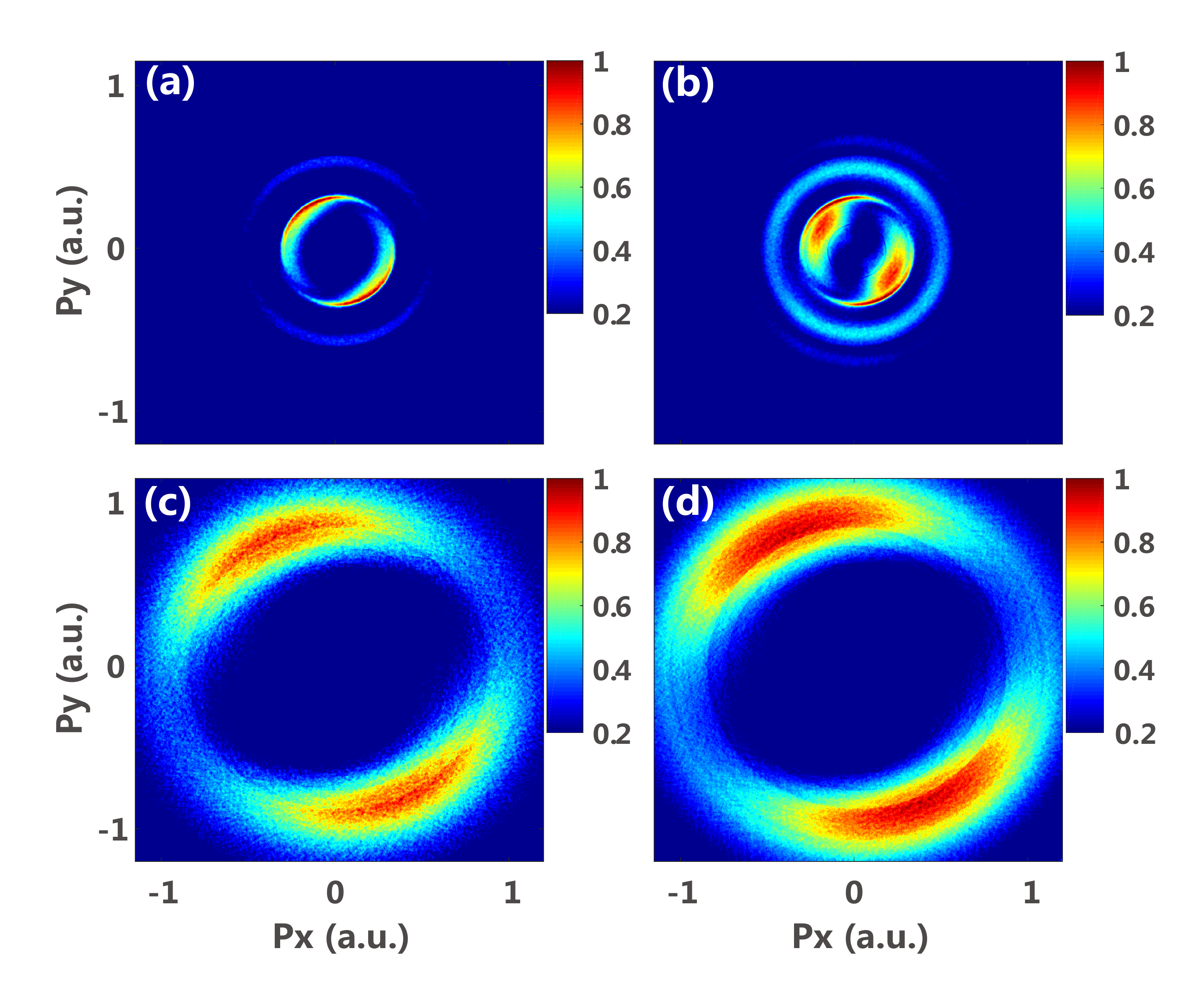}
  \caption{The measured 2D photoelectron momentum distributions in the polarization plane of argon ionized by near-circularly polarized fields at (a), (b) 410 nm and (c), (d) 800 nm. The large ellipticity is (a), (b) $\varepsilon = 0.853$ and (c), (d) $\varepsilon = 0.87$. The laser intensity is (a)$\thicksim0.08\times10^{15}W/cm^{2}$, (b)$\thicksim0.14\times10^{15}W/cm^{2}$, (c)$\thicksim0.19\times10^{15}W/cm^{2}$ and (d)$\thicksim0.23\times10^{15}W/cm^{2}$ .}
  \label{fig1}
\end{figure}

In Fig. 1, we show the reconstructed PMDs of argon in the polarization plane ionized by near-circularly polarized fields at 410 nm and 800 nm for different laser intensity. Here $p_{x}$ and $p_{y}$ are the momenta along the major and minor axes of the laser ellipse, respectively. As shown in Fig. 1(a) and (b), the ringlike ATI structures for 410 nm are easy to distinguish because of the large photon energy (0.114 a.u.). As for 800 nm, the valence electron needs to absorb more than eight photons. Therefore many excited states will contribute to the ionization and the ATI interference structures will be wash out as shown in Fig. 1(c) and (d). With increasing laser intensity, the ATI structures for 800 nm are found to shift towards higher energy. In the PMDs of 410 nm, however, the first-order ATI structure does not exhibit this energy shift due to the resonant ionization. Instead, this ringlike structure exhibits a remarkable spreading towards lower energy. Especially, referring to the unchanged first-order ATI peak, this spreading in Fig. 1(b) is more noticeable with higher laser intensity compared to that in Fig. 1(a), and exhibits a clear energy-dependent angular shift.

Intuitively, the spreading and angle offset shown in Fig. 1(b), similarly to the deflection of the outgoing electron observed in many theoretical and experimental studies [33-36], can be associated with the Coulomb interaction. However, the electron motion in the laser polarization plane is determined by the coupled laser and Coulomb field. In order to exclude the influence of the laser field and emphasize the role of the Coulomb effect, we therefore reconstruct the transverse momentum distributions ($p_{z}$) which is free of laser field. The transverse momentum distributions in near-circularly polarized field at 410 nm and 800 nm are plotted in Fig. 2(a) and (b), respectively. These distributions are given by the integration of the reconstructed three-dimensional electron wave packet over the $p_{x}$ and $p_{y}$. At first glance, the transverse momentum distributions by near-circularly polarized field at 410 nm is narrower than that at 800 nm. The transverse momentum distributions for 800 nm are Gaussian shape. As for 410 nm, the shape of the distributions exhibits a sharper peak. This phenomenon agrees well with the wavelength dependence of the transverse momentum distribution observed in previous works [27,28]. However, a closer examination reveals that, in contrast to the broadening of the distributions for 800 nm, the distributions for the 410-nm laser field are narrowed with the increase of the laser field. This narrowing violates the conventional rules that the transverse momentum distribution will broaden with increasing laser intensity (indicated by the ionization rate $P \propto \exp[-{2\left(2 I_{p}\right)^{2 / 3}}/{(3 F)}]$, here $I_{p}$ is the ionization potential, $F$ is the electric field strength) [47].

\begin{figure}
  \centering
  \includegraphics[width=8.5cm]{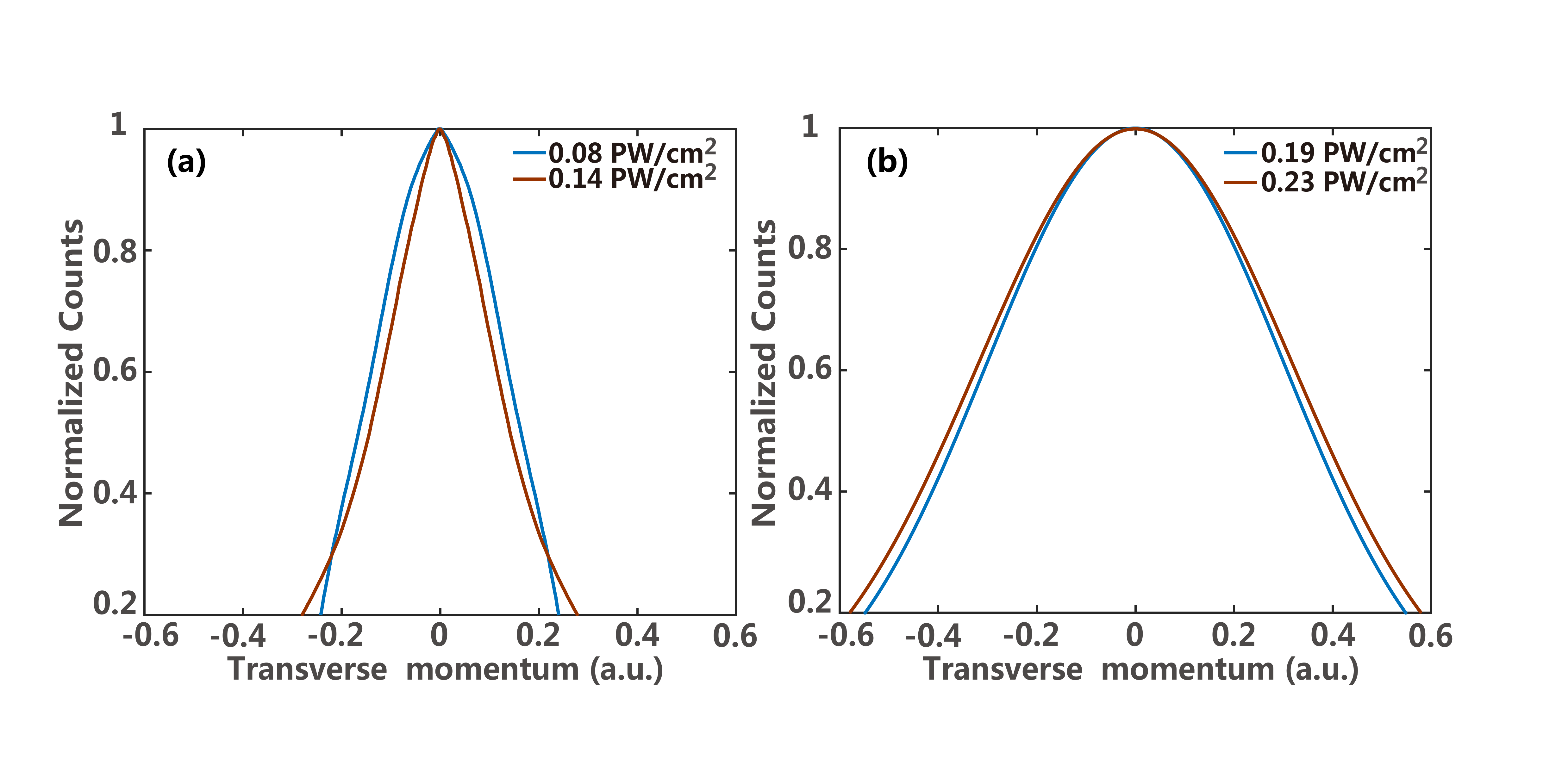}
  \caption{The measured transverse momentum distributions with the same laser intensity as in Fig.1 at (a) 410nm and (b) 800nm. The fitting curves are normalized to show the envelopes more clearly.}
  \label{fig2}
\end{figure}

In order to investigate the connection between this abnormal narrowing and Coulomb field, we employ the nonadiabatic CTMC method to simulate the electronic dynamics [48-51]. To describe the ionization process including the nonadiabatic effect, the initial momentum distribution is given by Perelomov-Popov-Terent¡¯ev (PPT) theories [48]. In this case, a momentum shift $\Delta p$ will be introduced into the initial momentum in the polarization plane [49,50]. This nonzero momentum shift can be expressed as
\begin{align}
\Delta p=\frac{\epsilon E(t)}{\omega}\left(\frac{\sinh \tau_{0}}{\tau_{0}}-1\right)
\end{align}
where $\epsilon$ is the ellipticity, $E(t_{0})$ is the electric field amplitude at the instant $t = t_{0}$, $\omega$ is the central frequency, and $\tau_{0}$ is obtained by
\begin{align}
\sinh ^{2} \tau_{0}\left[1-\epsilon^{2}\left(\operatorname{coth} \tau_{0}-\frac{1}{\tau_{0}}\right)^{2}\right]=\gamma^{2}
\end{align}
$\gamma$ is the Keldysh parameter ($\gamma=\sqrt{I_{p} / 2 U_{p}}$, $U_{p}$ is the ponderomotive energy). The tunneling exit $r_{0}$ after considering the nonadiabatic effect is calculated with [49]
\begin{align}
r_{0}=\frac{E(t)}{\omega^{2}}\left(\cosh \tau_{0}-1\right)
\end{align}
Then, the evolution of the electrons in the superposition field formed by the laser field and Coulomb field is governed by the classical
Newtonian equation, i.e. $d^{2} \vec {r}(t) / d t^{2} = -\nabla \cdot V(\mathbf{r}) -\vec {E}(t)$. Finally, the electrons having energies larger than 0 is regarded as the ionization events. The photoelectron momentum distribution can be produced by summing the electrons with the same momentum.

\begin{figure}
  \centering
  \includegraphics[width=8.5cm]{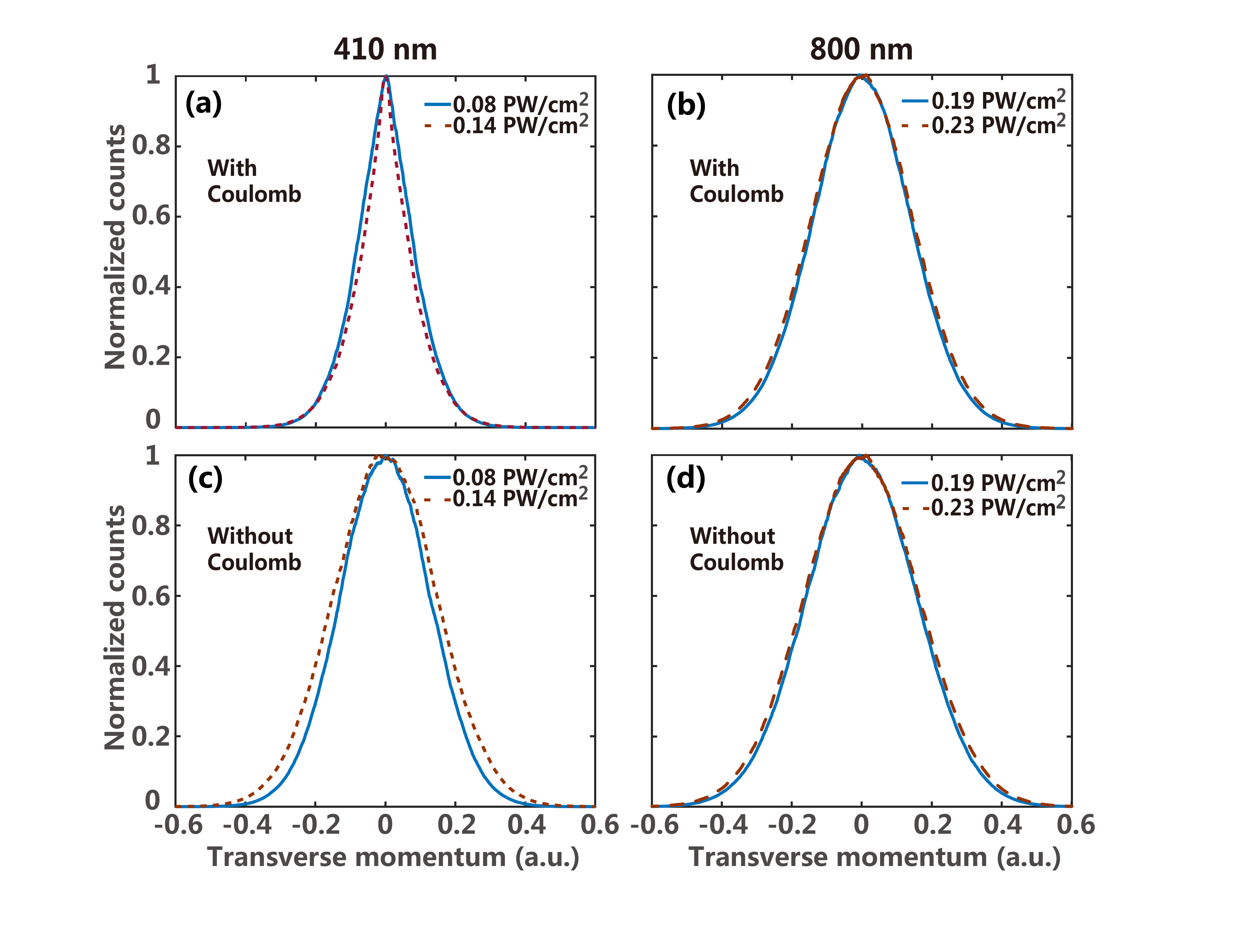}
  \caption{The simulated transverse momentum distributions in elliptically polarized light with (up) and without (down) inclusion of Coulomb interaction at 410 nm (left) and 800 nm (right). The solid lines present the calculations for lower intensity while the dashed lines for the higher.}
  \label{fig3}
\end{figure}

Figure 3(a) and (c) show the calculated transverse momentum distributions in a 410-nm wavelength field for different intensities with and without Coulomb potential, respectively. The ellipticity is the same as experiments ($\varepsilon = 0.853$). A $sin^{2}$ envelope is used with a duration of ten cycles. One can see the transverse momentum distribution with Coulomb interaction is much narrower than the distributions obtained without Coulomb interaction. The simulated distributions also exhibit a cusp-like shape, which is obviously out of Gaussian shape. In addition, with the increasing laser intensity, the distributions exhibit a stronger narrowing effect. When the Coulomb potential is neglected, however, the distributions are broadened as shown in Fig. 3(c). For comparison, we perform the simulations of the transverse momentum with and without Coulomb potential in an 800-nm wavelength field in the same parameter as the Fig. 2(b). As shown in Fig. 3(b) and (d), the Coulomb potential makes little difference to the transverse momentum Gaussian distributions. And the distributions are broadened as the increasing intensity, either with or without the Coulomb potential. The simulations in near-circularly polarized field at 410 nm and 800 nm both agree well with the experimental results in Fig.2. The theoretical simulations confirm that the Coulomb interaction is responsible for the narrowing of transverse momentum distributions presented in experiments, which is so-called Coulomb focusing.

The above results demonstrate the important role of Coulomb potential in the evolution of ejected electrons. However, the detailed electron dynamics under the coupled laser and Coulomb field remain to be solved. To this end, we now turn back to the momentum distribution in the polarization plane. The data in Fig. 1(b) is transformed into the angle-resolved photoelectron energy distributions to facilitate the analysis of the PMDs as shown in Fig. 4(a). It is clear to distinguish the spreading of the first-order ATI structure towards the lower energy. In addition, these lobes exhibit significant angular shift singularized by the white dashed curves. The lobes bend more seriously at the lower energy region, which indicates the electron with smaller final momentum has been subjected to stronger Coulomb effects. Moreover, Figure 4(b) demonstrates the transverse momentum distributions for different energy regions. Coincidentally, the curves responding to the lower energy region exhibit a narrower shape. The focusing in transverse momentum shows a good consistency with the spreading and angular shift in the polarization plane. To understand the physics underlying this phenomenon, we next analyze the motion of electrons with different energy.

Benefiting from the nonadiabatic CTMC model, we can trace the electron trajectory after ionizing. The time-dependent trajectories with different initial positions and momenta are simulated according to the classical Newtonian equation. The one with the maximum ionization rate is selected to stand for other trajectories with the same initial transverse momentum. To gain insight into the phenomenons in the low-energy region, we simulate the spatial trajectories corresponding to specific final kinetic energy in the polarization plane. Figure 4(c) shows the trajectory with the specific kinetic energy 1.14 eV. At first glance, this trajectory looks like a normal trajectory observed in the direct ionization, considering the rescattering is highly suppressed in a near-circularly polarized field. For a closer examination, the corresponding temporal evolution of the total energy of photoelectron, which is given by $\epsilon=\frac{1}{2} p^{2}-V(\mathbf{r})$ ($V(\mathbf{r})=\left(Z_{\mathrm{eff}} /|\mathbf{r}|\right)$ is the Coulomb potential and $Z_{\mathrm{eff}}$ is the effective nuclear charge), is plotted in Fig. 4(f). One can see that the electron is transiently trapped into a negative energy state. Before acquiring enough energy to be set free, the electron only can move around the nucleus as the trajectory in Fig. 4(d). Finally, the electron is set free with positive energy. It is worth mentioning that this process is different from FTI, in which the electron is also found to be trapped into negative energy [16-22], which will maintain in the Rydberg state rather than free into positive energy eventually. However, the quasibound state only serves as an intermediate state in our simulations.

\begin{figure}
  \centering
  \includegraphics[width=8.8cm]{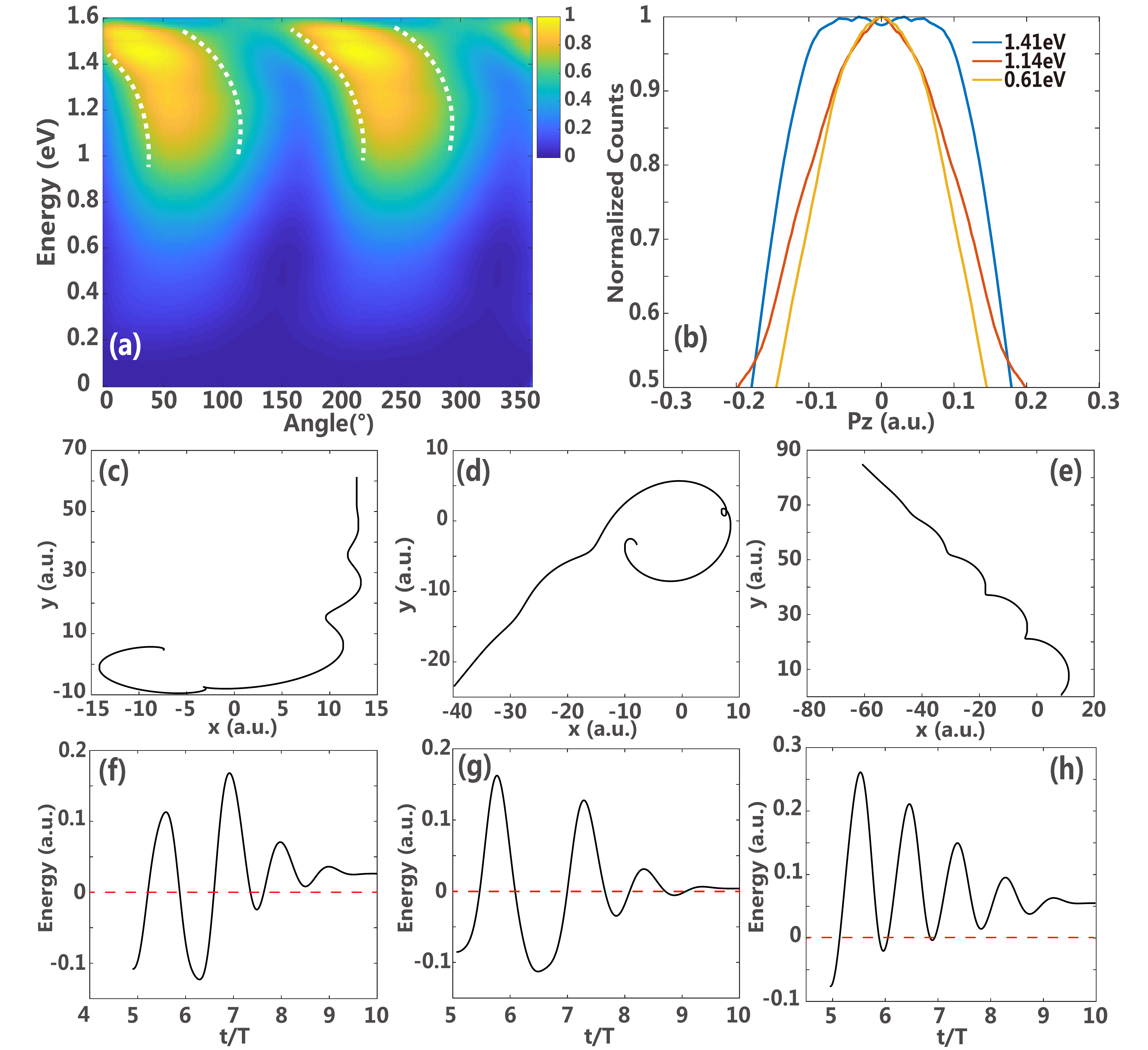}
  \caption{(a) The measured angle-resolved photoelectron energy distributions based on Fig. 1(b). The white dashed curves indicate the variation trend of the PMD as the electronic energy. (b) The measured transverse momentum distribution of the electron wave packet after ionizing for different energy regions. (c)-(e) The typical trajectories of electrons with the final kinetic energy $\thicksim$1.14 eV, $\thicksim$0.61 eV and $\thicksim$1.41 eV, respectively. (f)-(h) The corresponding temporal evolution of the total energy of the photoelectron shown in (c)-(e), respectively. The red dash line is plotted to distinguish the zero energy.}
  \label{fig4}
\end{figure}

The above observation indicates that the motion of the electron is bound up with its energy variation. Figure 4(d) and (g) illustrate a typical trajectory at the lower-energy region ($\thicksim$0.61 eV) and the corresponding temporal evolution of its energy. In this case, the residence time of the electron in the Rydberg state is much longer compared to that in Fig. 4(c) and (f). The electron thus experiences a stronger Coulomb effect and finally is closer to the core before the end of the laser pulse. Conversely, the electron at the higher-energy region ($\thicksim$1.41 eV, which is corresponding to the first-order ATI ring) only stays in the quasibound state for a shorter time (even not trapped into negative energy for the higher-energy region). Before it is finally emitted, the electron escapes away directly rather than circling the nucleus as shown in Fig. 4(e).

These typical trajectories demonstrate that the electron with lower energy is more likely to be retrapped into quasibound state. Before acquiring enough energy to be set free, the electron only can move around the nucleus as the trajectory in Fig. 5(d). Therefore the longer residence time in quasibound state means that the electron will be more strongly affected by the Coulomb potential within the laser cycles. The energy-dependent angular shift can also be attributed to this reason. And it is consistent with the characteristic of measured transverse momentum distribution as shown in Fig. 4(b). Moreover, in analogy to the narrowing in Fig. 2(b) and Fig. 3(a), the comparison between the PMDs in Fig. 1(a) and (b) demonstrates the enhancement of this effect produced by higher laser intensity. This enhancement can be explained by a raising of the ionization threshold produced by the ponderomotive energy ($U_{p} \propto I\lambda^{2}$, here $I$ is the laser intensity and $\lambda$ is the wavelength) shift [52]. This shift will reduce the initial kinetic energy of the emitted electron by decreasing its initial momentum in the polarization plane (i.e. $p_{y}$ and $p_{x}$), which will significantly contribute to the retrapped ionization and then strengthen the effect of Coulomb focusing. However, this effect makes no difference to the PMDs in the 800-nm near-circularly polarized field referring to Fig. 1(c) and (d). Instead, Figure 3(b), (c) and (d) exhibit a slight broadening due to the increase of ionization rate.

In conclusion, we have measured the three-dimensional PMDs of argon ionized by strong near-circularly polarized fields and reproduced the measurements based on nonadiabatic the CTMC. By disentangling photoemission pathways corresponding to different electron energy, we figure out that the emitted electron can be temporarily retrapped by atomic potential into a quasibound state. Before acquiring enough energy to be set free, the electron will be forced to circle the core and thus experience a stronger Coulomb effect. This physical mechanism leads to the significant focusing of first-order ATI structure towards the low-energy region in the polarization plane, and the narrowing transverse distributions with increasing laser intensity. Our findings provide a deep insight into electron dynamics in the Coulomb coupled system.

\section*{Funding}
National Natural Science Foundation of China (NSFC) (11574101, 11674116, 11934006, 11774111).

\section*{Disclosures}
The authors declare no conflicts of interest.


\begin{thebibliography}{50}
\bibitem{1} P. B. Corkum, "Plasma perspective on strong field multiphoton ionization," Phys. Rev. Lett. {\bf 71}, 1994 (1993); M. Hentschel, R. Kienberger, C. Spielmann, G. A. Reider, N. Milosevic, T. Brabec, P. Corkum, U. Heinzmann, M. Drescher, and F. Krausz, "Attosecond metrology," Nature (London) {\bf 414}, 509 (2001).
\bibitem{2} J. L. Krause, K. J. Schafer, and K. C. Kulander, "High-order harmonic generation from atoms and ions in the high intensity regime," Phys. Rev. Lett. {\bf 68}, 3535 (1992).
\bibitem{3} D. Shafir, H. Soifer, B. D. Bruner, M. Dagan, Y. Mairesse, S. Patchkovskii, M. Y. Ivanov, O. Smirnova, and N. Dudovich, "Resolving the time when an electron exits a tunnelling barrier," Nature (London) {\bf 485}, 343 (2012).
\bibitem{4} Q. Zhang, P. Lan, W. Hong, Q. Liao, Z. Yang, and P. Lu, "The effect of controlling laser field on broadband suppercontinuum generation," Acta Phys. Sin. {\bf 58}, 4908 (2009).
\bibitem{5} P. Agostini, F. Fabre, G. Mainfray, G. Petite, and N. K. Rahman, "Free-free transitions following six-photon ionization of xenon atoms," Phys. Rev. Lett. {\bf 42}, 1127 (1979).
\bibitem{6} D. N. Fittinghoff, P. R. Bolton, B. Chang, and K. C. Kulander, "Observation of nonsequential double ionization of helium with optical tunneling," Phys. Rev. Lett. {\bf 69}, 2642 (1992).
\bibitem{7} B. Walker, B. Sheehy, L. F. DiMauro, P. Agostini, K. J. Schafer, and K. C. Kulander, "Precision measurement of strong field double ionization of helium," Phys. Rev. Lett. {\bf 73}, 1227 (1994).
\bibitem{8} A. Becker and F. H. M. Faisal, "S-Matrix analysis of coincident measurement of two-electron energy distribution for double ionization of He in an intense laser field," Phys. Rev. Lett. {\bf 89}, 193003 (2002).
\bibitem{9} M. Weckenbrock, D. Zeidler, A. Staudte, Th. Weber, M. Sch\"{o}ffler, M. Meckel, S. Kammer, M. Smolarski, O. Jagutzki, V. R. Bhardwaj, D. M. Rayner, D. M. Villeneuve, P. B. Corkum, and R. D\"{o}rner, "Fully differential rates for femtosecond multiphoton double ionization of neon," Phys. Rev. Lett. {\bf 92}, 213002 (2004).
\bibitem{10} M. Lewenstein, Ph. Balcou, M. Yu. Ivanov, Anne L'Huillier, and P. B. Corkum, "Theory of high-harmonic generation by low-frequency laser fields," Phys. Rev. A, {\bf 49}, 2117 (1994).
\bibitem{11}L. V. Keldysh, "Ionization in the field of a strong electromagnetic wave," Sov. Phys. JETP {\bf 20}, 1307 (1965).
\bibitem{12} F. H. M. Faisal, "Multiple absorption of laser photons by atoms," J. Phys. B {\bf 6}, L89 (1973).
\bibitem{13} H. R. Reiss, "Effect of an intense electromagnetic field on a weakly bound system," Phys. Rev. A {\bf 22}, 1786 (1980).
\bibitem{14} R. Moshammer, J. Ullrich, B. Feuerstein, D. Fischer, A. Dorn, C. D. Schr\"{o}ter, J. R. Crespo Lopez-Urrutia, C. Hoehr, H. Rottke, C. Trump, M. Wittmann, G. Korn, and W. Sandner, "Rescattering of ultralow-energy electrons for single ionization of Ne in the tunneling regime," Phys. Rev. Lett. {\bf 91}, 113002 (2003).
\bibitem{15} F. H. M. Faisal and G. Schlegel, "Signatures of photon effect in the tunnel regime," J. Phys. B. {\bf 38}, L223 (2005).
\bibitem{16} T. Nubbemeyer, K. Gorling, A. Saenz, U. Eichmann, and W. Sandner, "Strong-field tunneling without ionization," Phys. Rev. Lett. {\bf 101}, 233001 (2008).
\bibitem{17} B. Wang, X. Li, P. Fu, J. Chen, and J. Liu, "Coulomb potential recapture effect in above-barrier ionization in laser pulses," Chin. Phys. Lett. {\bf 23}, 2729 (2006).
\bibitem{18} T. Nubbemeyer, K. Gorling, A. Saenz, U. Eichmann, and W. Sandner, "Strong-field tunneling without ionization," Phys. Rev. Lett. {\bf 101}, 233001 (2008).
\bibitem{19} U. Eichmann, T. Nubbemeyer, H. Rottke, and W. Sandner, "Acceleration of neutral atoms in strong short-pulse laser fields," Nature (London) {\bf 461}, 1261 (2009).
\bibitem{22} H. Lv, W. Zuo, L. Zhao, H. Xu, M. Jin, D. Ding, S. Hu, and J. Chen, "Comparative study on atomic and molecular Rydberg-state excitation in strong infrared laser fields," Phys. Rev. A {\bf 93}, 033415 (2016).
\bibitem{21} H. Zimmermann, S. Patchkovskii, M. Ivanov, and U. Eichmann, "Unified time and frequency picture of ultrafast atomic excitation in strong laser fields," Phys. Rev. Lett. {\bf 118}, 013003 (2017).
\bibitem{22} Y. Zhao, Y. zhou, J. Liang, Z. Zeng, Q. Ke, Y. Liu, M. Li, and P. Lu, "Frustrated tunneling ionization in the elliptically polarized strong laser fields," Opt. Express, {\bf 27}, 21689 (2019).
\bibitem{23} E. S. Shuman, R. R. Jones, and T. F. Gallagher, "Multiphoton assisted recombination," Phys. Rev. Lett. {\bf 101}, 263001 (2008).
\bibitem{24} C. Liu and K. Z. Hatsagortsyan, "Origin of unexpected low energy structure in photoelectron spectra induced by midinfrared strong laser fields," Phys. Rev. Lett. {\bf 105}, 113003 (2010).
\bibitem{25} T. M. Yan, S. V. Popruzhenko, M. J. J. Vrakking, and D. Bauer, "Low-energy structures in strong field ionization revealed by quantum orbits," Phys. Rev. Lett. {\bf 105}, 253002 (2010).
\bibitem{26} C. Liu and K. Z. Hatsagortsyan, "Wavelength and intensity dependence of multiple forward scattering of electrons at above-threshold ionization in mid-infrared strong laser fields," J. Phys. B {\bf 44}, 095402 (2011).
\bibitem{27} D. Comtois, D. Zeidler, H. P\'{e}pin, J. C. Kieffer, D. M. Villeneuve and P. B. Corkum, "Observation of Coulomb focusing in tunnelling ionization of noble gases," J. Phys. B {\bf 38}, 1923 (2005).
\bibitem{28} D. Shafir, H. Soifer, C. Vozzi, A. S. Johnson, A. Hartung, Z. Dube, D. M. Villeneuve, P. B. Corkum, N. Dudovich, and A. Staudte, "Trajectory-resolved coulomb focusing in tunnel ionization of atoms with intense, elliptically polarized laser pulses," Phys. Rev. Lett. {\bf 111}, 023005 (2013).
\bibitem{29} C. Liu and K. Hatsagortsyan, "Coulomb focusing in above-threshold ionization in elliptically polarized midinfrared strong laser fields," Phys. Rev. A {\bf 85}, 023413 (2012).
\bibitem{30} T. Brabec, M. Ivanov and P. B. Corkum, "Coulomb focusing in intense field atomic processes," Phys. Rev. A {\bf 54} R 2551 (1996)
\bibitem{31} G. L. Yudin and M. Ivanov, "Correlated multiphoton double ionization of helium: The role of nonadiabatic tunneling and singlet recollision," Phys. Rev. A {\bf 63} 033404 (2001).
\bibitem{32} A. Tong, Q. Li, X. MA, Y. Zhou, and P. Lu, "Internal collision induced strong-field nonsequential double ionization in molecules," Opt. Express, {\bf 27},  6415 (2019).
\bibitem{33} L. Torlina, F. Morales, J. Kaushal, I. Ivanov, A. Kheifets, A. Zielinski, A. Scrinzi, H. G. Muller, S. Sukiasyan, M. Ivanov, and O. Smirnova, "Interpreting attoclock measurements of tunnelling times," Nat. Phys. {\bf 11}, 503 (2015).
\bibitem{34} P. Eckle, M. Smolarski, P. Schlup, J. Biegert, A. Staudte, M. Sch\"{o}ffler, H. G. Muller, R. D\"{o}rner, and U. Keller, "Attosecond angular streaking," Nat. Phys. {\bf 4}, 565 (2008).
\bibitem{35} P. Eckle, A. N. Pfeiffer, C. Cirelli, A. Staudte, R. D\"{o}rner, H. G. Muller, M. B\"{u}ttiker, and U. Keller, "Attosecond ionization and tunneling delay time measurements in helium," Science {\bf 322}, 1525 (2008).
\bibitem{36} N. Camus, E. Yakaboylu, L. Fechner, M. Klaiber, M. Laux, Y. Mi, K. Z. Hatsagortsyan, T. Pfeifer, C. H. Keitel, and R. Moshammer, "Experimental evidence for quantum tunneling time," Phys. Rev. Lett. {\bf 119}, 023201 (2017).
\bibitem{37} A. N. Pfeiffer, C. Cirelli, M. Smolarski, D. Dimitrovski, M. Abu-samha, L. B. Madsen, and U. Keller, "Attoclock reveals natural coordinates of the laser-induced tunnelling current flow in atoms," Nat. Phys. {\bf 8}, 76 (2012).
\bibitem{38} M. Liu, Y. Shao, M. Han, P. Ge, Y. Deng, C. Wu, Q. Gong, Y. Liu, "Energy-and momentum-resolved photoelectron spin polarization in multiphoton ionization of Xe by circularly polarized fields," Phys. Rev. Lett. {\bf 120}, 043201 (2018).
\bibitem{39} D. T. Pierece, R. J. Celotta, G.-C Wang, W. N. Unertl, A. Galejs, C. E. Kuyatt, and S. R. Mielczarek, "The GaAs spin polarized electron source," Rev. Sci. Instrum. {\bf 51}, 478 (1980).
\bibitem{40} S. A. Wolf, D. D. Awschalom, R. A. Buhrman, J. M. Daughton, S. von Moln\'{a}r, M. L. Roukes, A. Y. Chtchelkanova, and D. M. Treger, "Spintronics: a spin-based electronics vision for the future," Science {\bf 294}, 1488 (2001).
\bibitem{41} A. Eppink and D. Parker, "Velocity map imaging of ions and electrons using electrostatic lenses: Application in photoelectron and photofragment ion imaging of molecular oxygen," Rev. Sci. Instrum. {\bf 68}, 3477 (1997).
\bibitem{42} Available at: [http://physics.nist.gov].
\bibitem{43} R. R. Freeman, P. H. Bucksbaum, H. Milchberg, S. Darack, D. Schumacher, and M. E. Geusic, "Above-threshold ionization with subpicosecond laser pulses," Phys. Rev. Lett. {\bf 59}, 1092 (1987).
\bibitem{44} R. R. Freeman and P. H. Bucksbaum, "Investigations of above-threshold ionization using subpicosecond laser pulses," J. Phys. B: At. Mol. Opt. Phys. {\bf 24}, 325 (1991).
\bibitem{45} A. Rudenko, K. Zrost, C. D. Schr\"{o}ter, V. L. B. Jesus, B. Feuerstein, R. Moshammer and J. Ullrich, "Resonant structures in the low-energy electron continuum for single ionization of atoms in the tunnelling regime," J. Phys. B: At. Mol. Opt. Phys. {\bf 37}, L407 (2004).
\bibitem{46} M. Wollenhaupt, M. Krug, J. K\"{o}hler, T. Bayer, C. SarpeTudoran, and T. Baumert, "Three-dimensional tomographic reconstruction of ultrashort free electron wave packets," Appl. Phys. B {\bf 95}, 647 (2009).
\bibitem{47} M. V. Ammosov, N. B. Delone, and V. P. Krainov, "Tunnel ionization of complex atoms and of atomic ions in an alternating electric field," Sov. Phys. JETP 64, 1191 (1986).
\bibitem{48} A. M. Perelomov and V. S. Popov, "Ioni zation of atoms in an alternating electrical field. III," Sov. Phys. JETP. {\bf 25}, 336 (1967).
\bibitem{49} J. Wang, F. He, "Tunneling ionization of neon atoms carrying different orbital angular momenta in strong laser fields," Phys. Rev. A. {\bf 95}, 043420 (2017).
\bibitem{50} Q. Zhang, G. Basnayake, A. Winney, Y. Lin, D. Debrah, S. K. Lee, and W. Li, "Orbital-resolved nonadiabatic tunneling ionization," Phys. Rev. A. {\bf 96}, 023422 (2017).
\bibitem{51} S. Luo, M. Li, W. Xie, K. Liu, Y. Feng, B. Du, Y. Zhou, and P. Lu, "Exit momentum and instantaneous ionization rate of nonadiabatic tunneling ionization in elliptically polarized laser fields," Phys. Rev. A. {\bf 99}, 053422 (2019).
\bibitem{52} G. Mainfray and C. Manus, "Multiphoton ionization of atoms," Rep. Prog. Phys. {\bf 54}, 1333 (1991).

\end{thebibliography}
\end{document}